\documentclass[aps,twocolumn,prl,showpacs]{revtex4}
\usepackage{graphicx}% include figures
\usepackage{amssymb} % AMS math symbol
\usepackage{url}     % handle url
\usepackage{bm}      % bold math
\begin{document}

\newcommand{\ohm}{\ensuremath{\,\Omega}}
\newcommand{\kohm}{\ensuremath{\,\mbox{k}\Omega}}
\newcommand{\cbg}{\ensuremath{C_{\mathrm{BG}}}}
\newcommand{\ctg}{\ensuremath{C_{\mathrm{TG}}}}
\newcommand{\vsd}{\ensuremath{V_{\mathrm{SD}}}}
\newcommand{\vbg}{\ensuremath{V_{\mathrm{BG}}}}
\newcommand{\vtg}{\ensuremath{V_\mathrm{TG}}}
\newcommand{\vbgo}{\ensuremath{V'_\mathrm{BG}}}
\newcommand{\vbgn}{\ensuremath{V^D_\mathrm{BG}}}
\newcommand{\vtgn}{\ensuremath{V^D_\mathrm{TG}}}
\newcommand{\nb}{\ensuremath{n_\mathrm{b}}}
\newcommand{\nt}{\ensuremath{n_\mathrm{t}}}
\newcommand{\ef}{\ensuremath{E_\mathrm{F}}}
\newcommand{\vbgbar}{\ensuremath{V_{\mathrm{BG}}- \vbgn}}
\newcommand{\vtgbar}{\ensuremath{V_\mathrm{TG}-\vtgn}}
%\title{Effective Transport Gap in Dual-gated Bilayer Graphene}
\title{Electronic Transport in Dual-gated Bilayer Graphene at Large Displacement Fields}
\author{Thiti  Taychatanapat$^{1}$}
\author{Pablo Jarillo-Herrero$^{2}$}
\email{pjarillo@mit.edu}
\affiliation{$^{1}$Department of Physics, Harvard University, Cambridge, MA 02138 USA}
\affiliation{$^{2}$Department of Physics, Massachusetts Institute of Technology, Cambridge, MA 02139 USA}
\date{\today}
\begin{abstract}
We study the electronic transport properties of dual-gated bilayer graphene devices. We focus on the regime of low temperatures and high electric displacement fields,  where we observe a clear exponential dependence of the resistance as a function of displacement field and density, accompanied by a strong non-linear behavior in the transport characteristics. The effective transport gap is typically two orders of magnitude smaller than the optical band gaps reported by infrared spectroscopy studies. Detailed temperature dependence measurements shed light on the different transport mechanisms in different temperature regimes.

\end{abstract}

% 72.20.-i = Transport processes in semiconductors and insulators
% 72.80.Vp = Graphene electronic transport
% 73.20.Hb = Impurities at surfaces and interfaces
% 73.22.Pr = Graphene electronic structure

\maketitle

%%%%%%%%%%%%%%%%%%%%%%%%%%%%%%%%%%%%%%%%%%%%%%%%%%%%%%%%%%%%%%%%%%%%%%%%%
%%%%%%%%%%%%%%%%%%%%%%%%%%%%%%%%%%%%%%%%%%%%%%%%%%%%%%%%%%%%%%%%%%%%%%%%%

The ability to electrostatically tune and deplete the charge density in two-dimensional electron gases enables the fabrication of basic mesoscopic devices, such as quantum point contacts or quantum dots, which enhance our understanding of electronic transport in nanostructures~\cite{Datta_Etransport}. Creating such electrically tunable nanostructures in monolayer graphene, a novel two-dimensional system~\cite{Novoselov}, is far more challenging due to its gapless nature. In this respect, Bernal-stacked bilayer graphene (BLG) is an interesting material, because of the possibility of opening a band gap by breaking the symmetry between the top and bottom graphene sheets~\cite{McCann_PhysRevB.74.161403,Castro_PhysRevLett.99.216802,Min_PhysRevB.75.155115}.

The low-energy band structure of free-standing BLG is gapless but in the presence of an on-site energy difference between the bottom and top layers a band gap develops. Different methods have been employed to induce a band gap including molecular doping, coupling to the substrate, and electric displacement field generated by gate electrodes~\cite{Ohta,Zhou_natureMat,Zhou_PhysRevLett.101.086402,Oostinga,xia_graphene_????,Zhang,Mak_PhysRevLett.102.256405}. However, the low-temperature ($\leq100\,$K) transport characteristics of dual-gated BLG devices do not exhibit the strong suppression of conductance expected given the large band gaps (up to $250\,$meV) measured by infrared spectroscopy~\cite{Oostinga,xia_graphene_????,Zhang,Mak_PhysRevLett.102.256405}. In addition, only very weak non-linearities were found in the current versus source-drain voltage ($I$-$\vsd$) characteristics~\cite{Oostinga}, in contrast with the strong non-linear behavior of typical semiconducting devices. A more complete study is needed to address the transport characteristics of gapped BLG devices as well as the role played by disorder.

%%%%%%%%%%%%%%%%%%%%%%      Figure 1     %%%%%%%%%%%%%%%%%%%%%%%%%%%%%%%%%%%%%%
%%%%%%%%%%%%%%%%%%%%%%%%%%%%%%%%%%%%%%%%%%%%%%%%%%%%%%%%%%%%%%%%%%%%%%%%%%%%%%%
\begin{figure}
\begin{center}
\includegraphics[width=2.95in]{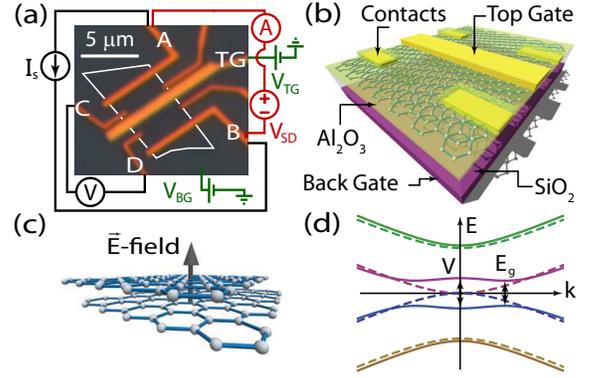}
\caption{(a) Optical image of a BLG (outlined by white line). A, B, C, and D are contact electrodes and TG is a $1\,\mu$m-wide top gate electrode. The red and black diagrams are set-ups for two and four probe measurements respectively. (b) Schematic diagram of the measured device (not drawn to scale). (c) When BLG is subject to a transverse electric field, a potential difference is induced between top and bottom layers. (d) Band structure of free standing BLG (dashed lines) and band structure of BLG subject to transverse electric field (solid lines).} \label{F:Device2}
\end{center}
%\vspace{-0.7cm}
\vspace{-0.5cm}
\end{figure}
%%%%%%%%%%%%%%%%%%%%%%%%%%%%%%%%%%%%%%%%%%%%%%%%%%%%%%%%%%%%%%%%%%%%%%%%%%%%%%%
%%%%%%%%%%%%%%%%%%%%%%%%%%%%%%%%%%%%%%%%%%%%%%%%%%%%%%%%%%%%%%%%%%%%%%%%%%%%%%%

In this letter, we report on the electronic transport properties of dual-gated (back-gated (BG) and top-gated (TG)) BLG devices. We focus on the regime of large transverse electric displacement fields, $0.8\,$V/nm $<\,|D|\,<\,2.5\,$V/nm, over 3 times larger than in previous low-temperature experiments ~\cite{Oostinga}. Upon the application of a large displacement field, we observe an exponential dependence of the device resistance on $|D|$ and density, which is accompanied by a strong non-linear behavior in the $I$-$\vsd$ characteristics. However, the size of the effective transport gap is on the order of a few meV, two orders of magnitude smaller than the optical band gaps at the same $D$~\cite{Zhang,Mak_PhysRevLett.102.256405}, suggesting a strong role played by disorder. Temperature dependent measurements in the $300\,$mK to $100\,$K range show that the conductivity follows an activated behavior with three different activation energies, including a nearest neighbor hoping regime at the lower temperatures. However, it is the conduction mechanism at intermediate temperatures ($2$-$70\,$K) which is responsible for most of the temperature variation of the conductivity of our devices, in which conductivity increases by several orders of magnitude.

BLG devices [Fig.~\ref{F:Device2}(a)] are fabricated by mechanical exfoliation~\cite{Novoselov_Pnas}, followed by the deposition of Cr/Au contact electrodes, Al$_2$O$_3$ growth, and deposition of a top gate~\cite{fabrication}. The typical mobility of our devices before oxide growth is between $1500\,$ and $2000\,$ cm$^2$/V$\cdot$s. The mobility can degrade significantly after oxide deposition (by $\sim$$30$\% in the device shown) which indicates that additional impurities have been introduced to the system during the oxide deposition. Charge and resonant impurities have been proposed to be dominant sources of scattering in graphene~\cite{Tan_minimumConductivity,Chen_chargeImpurity_MLG,Zhang_nature_physics,Xiao_ChargedBi,Ni_resonant_scatter}. The impurities alter the potential profile experienced by the charge carriers and lead to the formation of electron and hole puddles~\cite{Martin_nature_physics,Zhang_nature_physics,deshpande:243502}. They also induce tail states as well as localized states inside the band gap~\cite{Stauber_PhysRevB.76.205423,Johan_PhysRevLett.98.126801,Mkhitaryan_PhysRevB.78.195409,Wehling_dft_resonant}. In addition, in BLG devices, charged impurities could lead to a spatial variation of the band gap. This suggests that adsorbates and/or charged impurities in the oxide may play a significant role in electronic transport in BLG.

%%%%%%%%%%%%%%%%%%%%%%      Figure 2     %%%%%%%%%%%%%%%%%%%%%%%%%%%%%%%%%%%%%%
%%%%%%%%%%%%%%%%%%%%%%%%%%%%%%%%%%%%%%%%%%%%%%%%%%%%%%%%%%%%%%%%%%%%%%%%%%%%%%%
\begin{figure}
\begin{center}
\includegraphics[width=3.375in]{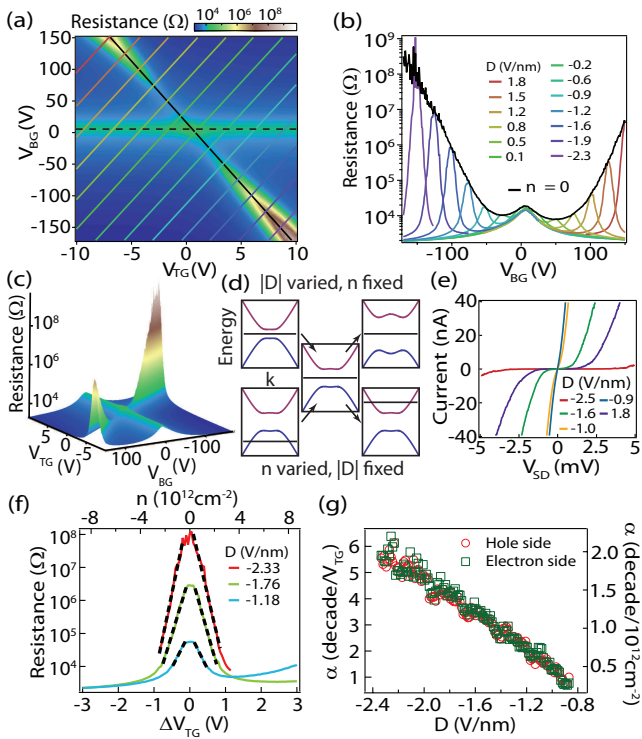}
\caption{(a) Differential resistance at zero bias as a function of top and back gate voltages at $T=300\,$mK (log scale). The black horizontal (dashed) and black diagonal (solid) lines correspond to zero charge densities in non top-gated and top-gated regions, respectively.  (b) Cuts in (a) at different displacement fields $D$ (colored lines) and at $n = 0$ (black line). Each cut corresponds to the lines in (a) with the same color. (c) 3D plot of (a). (d) Schematic band structure and $\ef$. As $D$ is varied at fixed $n$, the size of the band gap changes while $\ef$ remains fixed. We can also shift $\ef$ and keep the band gap constant by varying $n$ and keeping $D$ the same. (e) $I$-$\vsd$ characteristics at different values of $D$ for $n = 0$ at $T=300\,$mK. (f) The resistance as a function of carrier density and offset top gate voltages at various $D$'s. The black dashed lines are linear fits in log-scale. (g) The slope $\alpha$ of the linear fits in (f) as a function of $D$. The slope $\alpha$ decreases linearly with decreasing $D$.} \label{F:RvsVbgVtg}
\end{center}
%\vspace{-0.7cm}
\vspace{-0.4cm}
\end{figure}
%%%%%%%%%%%%%%%%%%%%%%%%%%%%%%%%%%%%%%%%%%%%%%%%%%%%%%%%%%%%%%%%%%%%%%%%%%%%%%%
%%%%%%%%%%%%%%%%%%%%%%%%%%%%%%%%%%%%%%%%%%%%%%%%%%%%%%%%%%%%%%%%%%%%%%%%%%%%%%%

We first focus on the transport properties at $300$~mK. Fig.~\ref{F:RvsVbgVtg}(a) shows the zero-bias resistance measured between electrodes A and B  in Fig.~\ref{F:Device2}(a).  A parallel plate capacitor model yields a charge density under the top-gated region $n = \cbg (\vbgbar) + \ctg (\vtgbar)$ where $C$ is the capacitive coupling, $V$ is the gate voltage, and $(\vbgn, \vtgn)$ is the charge neutrality point (CNP) in the top-gated region. Following the convention from Zhang~\emph{et al}~\cite{Zhang}, we define the average electric displacement field $D = (D_{\mathrm{BG}} + D_{\mathrm{TG}})/2$. In this letter, we concentrate on the case when $D_{\mathrm{BG}} = D_{\mathrm{TG}} = \epsilon_{\mathrm{BG}} ( \vbgbar/d_{\mathrm{BG}})$ where $\epsilon_{\mathrm{BG}} = 3.9$ is the relative dielectric constant of SiO$_2$ and $d_{\mathrm{BG}}=285\,$nm. The sharp rise in resistance along the $n=0$ [Fig.~\ref{F:RvsVbgVtg}(b)] is characteristic of BLG~\cite{Oostinga,xia_graphene_????,m._f._craciun_trilayer_2009}.The maximum on-off ratio we can achieve in this device at $300\,$mK is on the order of $10^6$, with a minimum resistance of $\sim$$300\,\Omega$ [measured in a four probe geometry at $(V_{\mathrm{BG}},V_{\mathrm{TG}}) = (-170,-10)\,$V]. We note that our high quality oxides enable us to apply a displacement field over $3$ times larger than in previous low temperature experiments~\cite{Oostinga}, resulting in an on-off ratio and insulating resistivity well over three orders of magnitude larger at $300\,$mK. Such insulating behavior makes BLG a good candidate for the fabrication of electrostatically designed mesoscopic devices.

The black diagonal curve in Fig.~\ref{F:RvsVbgVtg}(a-b) shows a slice of the resistance for $n = 0$. Along this curve,  we vary $D$ in the range $[-2.5,+1.8]\,$V/nm, while keeping $\ef$ in the top-gated region at the CNP [top path in Fig.~\ref{F:RvsVbgVtg}(d)]. Beyond $|D|\,\sim\,1\,$V/nm the resistance exhibits a clear exponential behavior with increasing $|D|$. Such resistance, if arising from a thermally activated behavior across a band gap, would be proportional to $\exp(E_g (D)/2k_BT)$ where $E_g(D)$ is the $D$-dependent band gap, $k_B$ is Boltzmann's constant, and $T$ is the temperature of the system. However, a fit to our data using the $E_g(D)$ obtained by infrared spectroscopy~\cite{Zhang}, yields an effective $T \approx 70\,$K, which is much higher than the $300\,$mK at which the measurement is performed. Hence, the conduction we observe cannot be explained by activated behavior across such an optical band gap: disorder plays an important role, and the associated energy scale for transport is about two orders of magnitude smaller. This exponential increase in resistance is also accompanied by the development of strong non-linear transport characteristics. Figure~\ref{F:RvsVbgVtg}(e) shows measurements of the DC current between A and B electrodes as a function of bias voltage and $D$, while keeping $n = 0$. We observe a clear non-linear behavior, which is consistent with the presence of an effective transport gap. However, the onset of non-linearity occurs on a scale of a few meV (2.2 meV for $|D|\,\sim\,2.5\,$V/nm), again about two orders of magnitude smaller than the optical band gap.

We now consider the behavior of the resistance versus gate voltage at various constant $D$ [Fig.~\ref{F:RvsVbgVtg}(b), colored lines]. By sweeping the gates along a constant $D$ line, we effectively hold the size of the band gap fixed and shift the Fermi energy $\ef$ from the valence band to the conduction band [bottom path in Fig.~\ref{F:RvsVbgVtg}(d)]. We observe an exponential decrease in resistance as we sweep the gate voltages away from the CNP ($n=0$), followed by a slower decrease, which we associate with $\ef$ reaching the valence and conduction band mobility edges. The decrease appears symmetric on both the electron and hole sides, and depends on $D$. Figure~\ref{F:RvsVbgVtg}(f) shows a few traces in detail, together with the fitted straight lines in the exponential regions, and Fig.~\ref{F:RvsVbgVtg}(g) shows the slope $\alpha$ of these lines as a function of $D$, which exhibitis an approximately linear behavior.

%%%%%%%%%%%%%%%%%%%%%%      Figure 3     %%%%%%%%%%%%%%%%%%%%%%%%%%%%%%%%%%%%%%
%%%%%%%%%%%%%%%%%%%%%%%%%%%%%%%%%%%%%%%%%%%%%%%%%%%%%%%%%%%%%%%%%%%%%%%%%%%%%%%
\begin{figure}
\begin{center}
\includegraphics[width=3.375in]{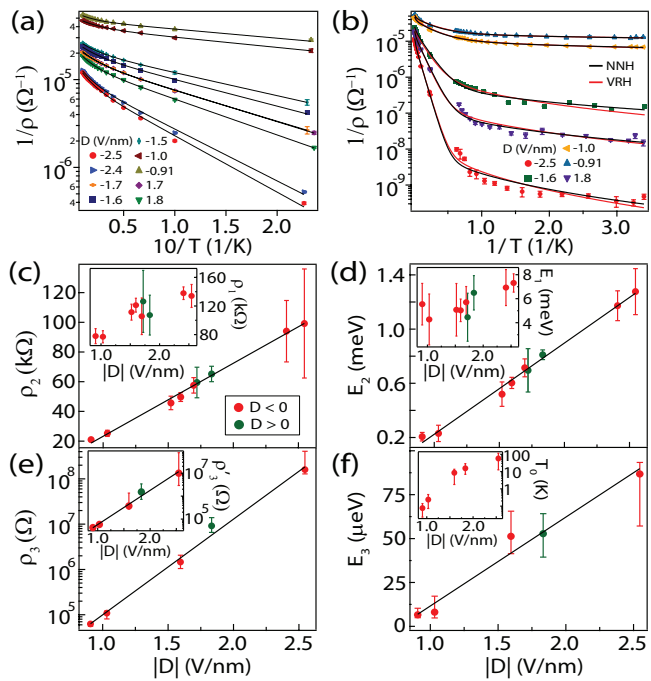}
\caption{(a) Conductivity as a function of inverse temperature from $4.2$ - $100$ K. (b) Conductivity as a function of inverse temperature from $300\,$mK to $100$ K. The black and red curves are the fits to the equation~\ref{E:R-Temp} with NNH and VRH terms respectively. (c)-(f) The extracted parameters from the fits plotted as a function of $|D|$.} \label{F:Tdependence2}
\end{center}
%\vspace{-0.7cm}
\vspace{-0.4cm}
\end{figure}
%%%%%%%%%%%%%%%%%%%%%%%%%%    Figure 3     %%%%%%%%%%%%%%%%%%%%%%%%%%%%%%%%%%%%
%%%%%%%%%%%%%%%%%%%%%%%%%%%%%%%%%%%%%%%%%%%%%%%%%%%%%%%%%%%%%%%%%%%%%%%%%%%%%%%

To gain further insight into the transport properties, we study the temperature dependence of the conductivity in the $300\,$mK to $100\,$K range at different $D$'s, and at $n = 0$. Figure~\ref{F:Tdependence2}(a) shows the conductivity measured using a four probe geometry in the temperature range $4$-$100\,$K for nine different values of $D$, while Fig.~\ref{F:Tdependence2}(b) shows the conductivity over the entire temperature range for five different values of $D$. In contrast to previous low temperature experiments at low $D$-fields~\cite{Oostinga}, we observe two distinct temperature regimes, which become more pronounced at high $D$: a two-component fast decrease of the conductivity from $100\,$ to $2\,$K, followed by a relatively weak $T$-dependence from $2\,$K to $300\,$mK, which we associate with thermally activated hopping (see below).

At intermediate to high temperatures, we find that the conductivity is well described by an activated behavior with two different activation energies, $1/\rho = \sum_{i=1}^{2} 1/\rho_i\exp(E_i/k_BT)$ where $\rho$ and $E$ are resistivity and activation energy respectively [black lines in Fig.~\ref{F:Tdependence2}(a)]. The extracted fit parameters ($\rho_1,\rho_2,E_1,$ and $E_2$) are plotted in Fig.~\ref{F:Tdependence2}(c-d). The higher energy scale, $E_1$, may be related to thermal activation across the optical band gap~\cite{xia_graphene_????}. However, due to the very limited high-$T$ range (70-100K), our data barely probe the onset of this exponential increase, and the values of $E_1$ obtained are very likely underestimated. The intermediate energy scale, $E_2$, exhibits a linear dependence on $D$, and is therefore approximately proportional to the band gap~\cite{Zhang}. However, its value is two orders of magnitude smaller than the observed optical band gap at the same $|D|$~\cite{Zhang}. Still, it is the conduction mechanism in this intermediate $T$-regime that is responsible for most of the measured variation in conductivity of the devices.

The overall behavior of the conductivity with $T$ is reminiscent of that observed in disordered semiconductors~\cite{Shklovskii}, where transport via impurity bands and thermally activated hopping dominate transport at low temperatures. However, BLG devices are unique, in that their band gap can be continuously tuned by electrostatic means, and also because its two-dimensionality means that all disorder effects are surface rather than bulk effects. Compressibility measurements~\cite{Henriksen_compressBi,Young_compressBi} have shown that a very large density of states exists in gapped BLG on SiO$_2$, even in moderate $D$ fields. Most of these states, however, are localized and do not contribute to transport, and in general the relationship between density of states and conductivity is more complex in the insulating regime.

To explain the origin and some of the qualitative features of the conductivity at intermediate $T$, we consider a model of an impurity band arising from the interaction of negatively charged donors~\cite{Nishimura_PhysRev.138.A815}. Such model has been used to explain the observed conduction in the intermediate temperature regime in germanium semiconductors~\cite{Fritzsche_PhysRev.125.1552,Davis_PhysRev.140.A2183}. Disorder can lead to the formation of an impurity band in which charge carriers are localized and a hopping mechanism dominates the conduction as we will show below. At low temperature, the majority of these localized states are empty or singly occupied. However, some of the states can become doubly occupied which leads to an extra band at higher energy due to the Coulomb interaction. Carriers in this band are weakly localized and hence possess a higher mobility. We can estimate the localization length of these states by equating $E_2$ with a charging energy $e^2/\epsilon r$ where $\epsilon \sim 5$ is the average dielectric constant of our top and bottom oxides, and $r$ is the localization length. This estimate yields a localization length on the order of $100\,$nm (at $|D| = 2.5\,$V/nm) to $1\,\mu$m at ($|D| = 0.9\,$V/nm), comparable with the width of the top gate. Such large localization length supports that the states are weakly bound. While this model accounts for some of the features present in the data, a rigorous theoretical model which takes into account the geometry and particularities of disorder in BLG, beyond the scope of this paper, is needed for a direct comparison and understanding of the complex relationship between transport and density of states, specially in the non-linear regime. In addition, compressibility measurements at larger $D$ and lower temperatures~\cite{Henriksen_compressBi,Young_compressBi} may have enough resolution to observe the effects of disorder at these energy scales.

The conductivity between $2\,$K and $300\,$mK decreases weakly with temperature, which indicates that we enter a hopping conduction regime through strongly localized states~\cite{Shklovskii}. We perform a fit to the conductivity for the complete temperature range and all $D$ with
\begin{equation}
    \frac{1}{\rho} = \frac{1}{\rho_1 \exp(\frac{E_1}{k_B T})} + \frac{1}{\rho_2 \exp(\frac{E_2}{k_B T})} + \frac{1}{\Xi},  \label{E:R-Temp}
\end{equation}
where $\Xi = \rho_3\exp(E_3/k_BT)$ for nearest neighbor hopping (NNH) or $\rho'_3\exp(T_0/T)^{1/3}$ for variable range hopping (VRH)~\cite{Shklovskii}. Both the NNH and VRH fits agree reasonably well with the data [Fig.~\ref{F:Tdependence2}(b)]. VRH has been proposed to be the transport mechanism for gapped BLG at intermediate and low $T$~\cite{Oostinga}. However our measurements yield a value of $\rho'_3$ that grows exponentially with $|D|$ [Fig.~\ref{F:Tdependence2}(e) inset]. Such strong dependence is unexpected because, in VRH, the factor $\exp(T_0/T)^{1/3}$ already includes the strong exponential contributions from both hopping between sites and differences in energy levels~\cite{Shklovskii}. Hence, we propose that the transport mechanism in our BLG devices is NNH in the temperature and $D$ regime explored.

For NNH, $\rho_3 = \rho_3^0 \exp(2r/a)$ where $r$ is the distance between hopping sites and $a$ is the localization length~\cite{Shklovskii}. The magnitude of $r$ can be approximated from the density of impurities, $n_{i}$, by $r = n_i^{-1/2}$. The CNP in this device is located at $(\vbgn, \vtgn) \approx (16,0)\,$V, which corresponds to $n_i \approx 10^{12}\,$cm$^{-2}$ and $r \approx 10\,$nm. The linear fit from Fig.~\ref{F:Tdependence2}(e) yields a localization length $a \approx 4/[D$(V/nm)$]\,$nm.  This allows us to estimate the crossover temperature $T^{\text{VRH}}$ at which the conduction mechanism changes from NNH at high $T$ to VRH at low $T$. As the temperature is lowered, it is feasible for electrons to hop to further sites but closer in energy due to the reduced coulomb interaction.  This transition takes place when $2r/a$ and $E_3/k_BT$ are comparable~\cite{Shklovskii}, which yields $T^{\text{VRH}} \lesssim 80\,$mK for $|D| < 2.5\,$V/nm. This crossover temperature is almost $4$ times smaller than the lowest $T$ we have studied ($300\,$mK) which supports that NNH dominates the conduction.

In addition, the hopping activation energy $E_3$ decreases with decreasing $|D|$ [Fig.~\ref{F:Tdependence2}(f)]. This is an indication of the BLG making a transition from a strong to a weakly insulating state. As the band gap gets smaller, the electron-hole puddles start to merge and hence create a channel for carriers to percolate from one electrode to the other~\cite{Adam_PhysRevLett.101.046404}. Extrapolating the fit to $E_3 = 0$, we obtain a value for the displacement field corresponding to this transition of $|D| = 0.7\,$V/nm corresponding to $E_g/2 \approx 35\,$meV which is consistent with BLG $\ef$ fluctuations for an impurity density $n_i \approx 10^{12}\,$cm$^{-2}$.

This work has been supported by ONR-MURI, the NSF-funded MIT CMSE and Harvard CNS. We thank M. Zaffalon for experimental help and A. H. Castro Neto, S. Das Sarma, M. A. Kastner, P. Kim, E. I. Rashba, A. Yacoby, and A. F. Young for discussions.

%\bibliography{TransportGapinBilayer}

\end{document}